\documentclass{acmart}

\usepackage{braket}

\AtBeginDocument{%
  \providecommand\BibTeX{{%
    \normalfont B\kern-0.5em{\scshape i\kern-0.25em b}\kern-0.8em\TeX}}}

\copyrightyear{2023}
\acmYear{2023}
\setcopyright{acmlicensed}\acmConference[NANOARCH '23]{18th ACM International Symposium on Nanoscale Architectures}{December 18--20, 2023}{Dresden, Germany}
\acmBooktitle{18th ACM International Symposium on Nanoscale Architectures (NANOARCH '23), December 18--20, 2023, Dresden, Germany}
\acmPrice{15.00}
\acmDOI{10.1145/3611315.3633259}
\acmISBN{979-8-4007-0325-6/23/12}

\begin{document}

\title{Towards Faster Reinforcement Learning of Quantum Circuit Optimization: Exponential Reward Functions}

\author{Ioana Moflic, Alexandru Paler}
\affiliation{%
  \institution{Aalto University}
  \country{Finland}
}
\email{alexandru.paler@aalto.fi}

\renewcommand{\shortauthors}{Moflic and Paler}
\renewcommand{\shorttitle}{Exponential Reward Functions}

\newcommand{\zero}{\textbf{\texttt{0}} }
\newcommand{\one}{\textbf{\texttt{1}} }

\begin{abstract}          
Reinforcement learning for the optimization of quantum circuits uses an agent whose goal is to maximize the value of a reward function that decides what is correct and what is wrong during the exploration of the search space. It is an open problem how to formulate reward functions that lead to fast and efficient learning. We propose an exponential reward function which is sensitive to structural properties of the circuit. We benchmark our function on circuits with known optimal depths, and conclude that our function is reducing the learning time and improves the optimization. Our results are a next step towards fast, large scale optimization of quantum circuits.
\end{abstract}


\maketitle

\setlength{\belowcaptionskip}{-10pt}

\section{Introduction}

There is a gap between the speed of the current generation of quantum circuit compilers and the performance required to handle large circuits. This performance gap has not been obvious until recently, because real quantum computers could only execute circuits of up to a few tens of qubits. Quantum circuit optimization using template-based rewrite rules \cite{miller2003transformation} is widely used in quantum circuit optimization tools. An input circuit is gradually transformed by applying quantum gate identities (e.g. Fig.~\ref{fig:circs}) until a given optimization criteria is met. The gate set and the input circuit size influence the performance of this approach. The number of allowed transformations increases the size of the optimization search space, thus scaling the procedure is a challenging task.

Machine learning techniques are being applied to quantum circuit compilation and optimization (e.g. \cite{paler2020machine, fan2022optimizing}). Quantum circuit optimization can be framed as a RL problem~\cite{fosel2021quantum, moflic2023graph, van2023qgym, li2023quarl}. Given the quantum circuit and a range of local rewrite rules to be applied, an agent will learn an optimal policy in a trial and error approach. The circuit represents a fully observable environment and applicable rewrites at a given time step are the actions. Both the RL state space and the action space are discrete, with the latter being size-dependent on the structure of the circuit. 

The goal of the \emph{reward function} is to maximize the agent's long term reward in order to solve the given problem. Herein we improve the scalability of Reinforcement Learning (RL) for quantum circuit optimization as presented by~\cite{fosel2021quantum}.

\section{Methods}
\label{sec:met}

\begin{figure}[!t]
    \centering
    \includegraphics[width=0.8\columnwidth]{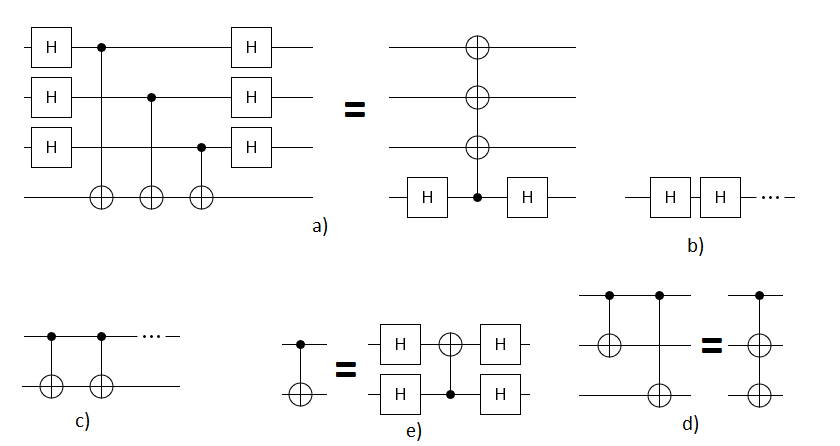}
    \caption{Quantum circuits can be optimized through by learning a heuristic that is applying template-based rewrite rules. In the process of reinforcement learning, an agent will learn how to apply the template rewrites. In this work we optimize Bernstein Vazirani circuits by applying four templates:  a) unoptimized and optimized Bernstein-Vazirani circuit; b) two Hadamard gates cancelling; c) two CNOT gates cancelling; d) parallelizing CNOTs sharing the same control; d) reversing the CNOT direction using Hadamard gates.}
    \label{fig:circs}
\end{figure}

We implement an RL method for circuit optimization and use Q-Learning to train our agent. The agent chooses between two actions encoded as \zero (the circuit transformation should not be performed) and \one (the circuit transformation should be performed). We use the templates from Fig.~\ref{fig:circs}b-e to optimize Bernstein-Vazirani circuits (Fig.~\ref{fig:circs}a) which have a known optimal depth of three (Fig.~\ref{fig:circs}a).

The literature on RL reward functions for quantum circuit optimization is sparse. For example, the work of~\cite{fosel2021quantum} uses a circuit cost function for expressing the reward: $r_t =-(q(s_{t+1}) - q(s_t))$, where $s$ is a state (\emph{i.e.} a quantum circuit), and $q(s)=len(s) - 0.2 count(s)$. The $len$ function returns the depth of the circuit and $count$ returns the gate count. The corresponding reward function is $r_t= (len(s_t) - len(s_{t+1}) - 0.2 (count(s_t) - count(s_{t+1}))$. The $r_t$ reward function increases when the depth is reduced, while \emph{decreasing gate counts would have a negative impact on the reward}. This reward function prefers circuit slices with increased gate parallelism. Another circuit cost function has been presented by~\cite{paler2020machine}: $Ratio(a, b) = \frac{len(a)}{len(b)}$, where $a$ and $b$ are two circuits.

We propose a novel reward function $R_{pow}$ as an exponential function of $Ratio$: a) it increases whenever the circuit becomes compatible with lower connectivity devices; b) it has a steeper increase whenever the depth of the circuit is reduced.
\begin{align*}
R_{pow}             &= cost + (str(s_t) - str(s_{t+1}))^{Ratio (s_t, s_{t+1})}\\
cost                &= \begin{cases}
                        0 & \text{if action is \zero}\\
                        c \in (0,0.2] & \text{if action is \one}
                    \end{cases}
\end{align*}

In the above equations, $str$ returns the interaction strength existing in a circuit. We define the \emph{interaction strength} $str(q1,q2)$ between two circuit wires $q1$ and $q2$ as the number of two qubit gates applied to the qubits $q1$ and $q2$ during the entire execution of the circuit. The \emph{interaction strength} $str(s)$ of a circuit $s$ is the mean of all the interaction strengths between the pairs of qubits in the circuit $s$. $R_{pow}$ has a positive value whenever the \emph{interaction strength} decreases, irrespective of how the depth of the circuit behaves. Depth improvements increase the reward exponentially, while depth increases have a sub-unit effect on the reward. 

The $R_{pow}$ function is preferring circuits with low interaction strengths. This is to make the function compatible with quantum chip layouts that have reduced connectivity: low depth, highly parallel circuits are not necessarily useful if the underlying quantum hardware has reduced connectivity between the qubits. At the same time, as illustrated Fig.~\ref{fig:circs}d, we allow gate parallelism in the form of single control multiple target CNOTs. As long as the mean degree is lowered, we assume that two-qubit gate parallelism does not impact the feasibility of the circuit's execution.

\section{Preliminary Results and Conclusion}
\label{sec:res}

We compare our $R_{pow}$ reward with the $Ratio$ reward~\cite{paler2020machine}. We focus explicitly on the optimization of stabilizer circuits consisting of CNOT and Hadamard gates. The comparison is based on how many epochs it takes to learn the optimization of a \emph{single circuit} and how deep the output circuit is. To quantify the performance we use circuits with known optimum depths (Fig.~\ref{fig:circs}a): learning takes the number of epochs until the optimum is reached. The agent applies rewrite rules with the goal of learning the best strategy.

\begin{table}[!h]
    \renewcommand{\arraystretch}{0.7}%
    \small
    \centering
    \begin{tabular}{c|c|c|c|c|c|c}
qubits	&min(rew)	&max(rew)	&mind	&fq(mind)	&maxd	&fq(maxd)\\
\hline
3&	2.25&	4.20&	3&	296&	8&	78\\
6&	5.17&	7.63&	3&	9&	14&	12\\
9&	8.2&	10.24&	6&	7&	20&	2\\
12&	11.09&	13.27&	7&  5&	25&	2\\
\hline
3&	    0&	    1.91&	3&	6829&	8&	80\\
6&	    0&	    2.55&	3&	6181&	14&	12\\
9&	    0.2&	3.19&	3&	5763&	20&	2\\
12&	    0.4&	3.83&	3&	5519&	24&	3
\vspace{10pt}
    \end{tabular}
    \caption{Comparing $Ratio$ (existing -- the first four lines) vs $R_{pow}$ (our proposal -- the last four lines) as reward functions. The \emph{min(rew)} and \emph{max(rew)} are minimum and maximum reward during one of the 8000 epochs. The \emph{mind} and \emph{maxd} refer to the min and max depths, and the $fq$ columns is the frequency of those values in the 8000 epochs.}
\label{tab:comparison}
\end{table}

\begin{figure}[!t]
    \centering
    \includegraphics[width=0.7\columnwidth]{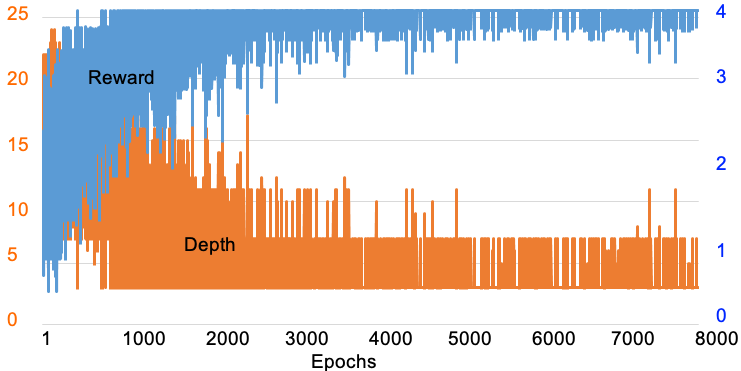}
    \caption{The $R_{pow}$ function does perform well when used as a reward function. These plots are for the 12 qubit Berstein-Vazirani circuit whose optimum depth is 3. The reward increases simultaneously with the decreasing optimum depth for the first approx. 3000 epochs. The value of reward refers to the \emph{cumulative reward per epoch} (the sum of all the per step rewards).}
    \label{fig:pow12}
\end{figure}

We evaluate the time it takes to train an agent for 8000 epochs and the achieved circuit depth. Table~\ref{tab:comparison} synthesizes the results of using both reward functions. Four types of Bernstein Vazirani circuits were analyzed having 3,6,9,12 qubits each. The $R_{pow}$ function performs significantly better. The minimum depth of three is achieved on each circuit instance. The frequency of finding the minimum depth is an indication of the learning speed: it takes approx. $8000 - freq(mind)$ epochs to find the minimum depth. The cumulative reward per epoch increases with the circuit size. Fig.~\ref{fig:pow12} illustrates how our RL framework performs on the 12 qubit Bernstein-Vazirani circuit when it uses $R_{pow}$.

\begin{acks}
This research was developed in part with funding from the Defense Advanced Research Projects Agency [under the Quantum Benchmarking (QB) program under award no. HR00112230007 and HR001121S0026 contracts]. We acknowledge the funding received from the Finnish-American Research and Innovation Accelerator. We thank Niki Loppi of the NVIDIA AI Technology Center Finland for his help with the implementation.
\end{acks}

\bibliographystyle{ACM-Reference-Format}
\bibliography{__paper}

\end{document}